\documentclass[prd,showpacs,amsmath,amssymb]{revtex4-2}
\usepackage{amsmath}
\usepackage{amsfonts}
\usepackage{mathrsfs}
\usepackage{pstricks}
\usepackage{color}
\usepackage{graphicx}
\usepackage{slashed}
\usepackage{amssymb}
\usepackage{hyperref}

\newcommand{\be}{\begin{equation}}
\newcommand{\ee}{\end{equation}}
\newcommand{\ba}{\begin{array}{c}}
\newcommand{\ea}{\end{array}}
\newcommand{\bqa}{\begin{eqnarray}}
\newcommand{\eqa}{\end{eqnarray}}
\makeatletter
    
    \newcommand{\Rmnum}[1]{\expandafter\@slowromancap\romannumeral #1@}
    \makeatother

\begin{document}

\title{\bf Compositeness relations for near-threshold p-wave bound states}

\author{Guo-Ying Chen}

\affiliation{Department of Physics and Astronomy, Hubei University
of Education, Wuhan 430205, China}


\date{\today}

\begin{abstract}
We generalize Weinberg's compositeness relations to near-threshold p-wave bound states and derive the relations among the p-wave
effective range expansion parameters, binding energy B, and the
field renormalization constant Z. Unlike the s-wave case, the field renormalization constant Z for p-wave states depends on the renormalization scale. We obtain the renormalization equation for Z and provide its solution. We find that the negative-norm state reported more than a decade ago in p-wave sector can be avoided by an appropriate choice of the renormalization scale. We also present the general propagator for p-wave near-threshold states. Our results are expected to have broad applications in studying exotic states, as well as the structure of halo nuclei and Fashbach molecules.
\end{abstract}


\maketitle

\section{Introduction}
Recently, the BESIII Collaboration reported a structure denoted as
G(3900)~\cite{BESIII:2024ths,Wang:2025dur} in the inelastic electron-positron scattering into $D^0\bar D^0$ and $D^+D^-$ final states. This structure was also
reported previously by BaBar~\cite{BaBar:2006qlj} and
Belle~\cite{Belle:2007woe}. The discovery of G(3900) sparked
interest in p-wave molecular
states~\cite{Chen:2025gxe,Lin:2024qcq,Premoli:2026lrb,Nakamura:2023obk,Ye:2025ywy}. Besides the molecular interpretation, it is also desirable to develop a theoretical framework which can systematically distinguish whether a near-threshold p-wave state is a composite molecular state or an elementary state, such as a tetraquark state (see, e.g.,~\cite{Lodha:2024bwn,lodha:2025snn}). Such a theoretical framework is important not only for studying the nature of p-wave near-threshold exotic states in hadron physics but also for the study of p-wave halo nuclei in nuclear physics and p-wave Feshbach molecules in ultracold atom physics~\cite{RevModPhys.76.215,PhysRevLett.98.200403,PhysRevLett.101.100401,smgh-m2yq}.

In the 1960s, Weinberg proposed the compositeness relations to
distinguish a composite state from an elementary state for
near-threshold s-wave states\cite{Weinberg:1962hj,Weinberg:1965zz}.
These relations are
\begin{equation}
a=\frac{2(1-Z)}{2-Z}\frac{1}{\sqrt{2\mu B}},\ \ \ \ \ \
r_0=-\frac{Z}{1-Z}\frac{1}{\sqrt{2\mu B}},\label{ar}
\end{equation}
where $a$ is the scattering length, $r_0$ is the effective range,
$Z$ is the field renormalization constant of the bound state, $\mu$
is the reduced mass of the two-particle system, and B is the binding
energy. A near-threshold state is identified as a pure molecular state if $Z=0$, while a non-zero 
$Z$ implies the presence of an elementary component. In recent years, the concept of compositeness has attracted
renewed
interest\cite{Baru:2003qq,Aceti:2012dd,Chen:2013upa,Sekihara:2014kya,Guo:2015daa,vanKolck:2022lqz},
since many exotic states have been reported over the past two
decades. Although several generalizations of the Weinberg compositeness condition to arbitrary partial waves have been discussed in the literature~\cite{Aceti:2012dd,Kinugawa:2024crb,Bruns:2025xgo}, the closed-form near-threshold p-wave relations that directly connect the effective range expansion parameters to Z and B have not been explicitly presented.
The aim of this manuscript is to generalize Weinberg's
compositeness relations to p-wave bound states within the
nonrelativistic effective field theory (NREFT) for hadrons, originally proposed by Weinberg~\cite{Weinberg:1990rz,Weinberg:1991um,Weinberg:1992yk}. The framework reported in the manuscript differs from the halo effective field theory for p-wave near-threshold states~\cite{Bertulani:2002sz,Bedaque:2003wa,Rupak:2011nk,Hammer:2011ye}, as we explicitly include the factor $Z$ in the formulas, making the analysis of non-vanishing $Z$ cases more straightforward.

This paper is organized as follows. In Sec.~\ref{swave}, we propose a
more direct approach to derive the compositeness relations for s-wave bound
states within the framework of NREFT. This approach is then used to
obtain the compositeness relations for p-wave bound states in
Sec.~\ref{pwave}. In Sec.~\ref{negative}, we resolve the problem of the negative norm state, which was pointed out more than a decade ago, and demonstrate that such a state can be avoided by an appropriate choice of the renormalization scale.
In Sec.~\ref{Feynmanrules}, we present the general propagator for p-wave near-threshold states, which can be used to fit the experimental lineshape data. Finally, we summarize our results in Sec.~\ref{summary}.

\section{S-wave compositeness relations via on-shell renormalization in NREFT}\label{swave}

\begin{figure}[t]
 \begin{center}
  \includegraphics[width=10cm]{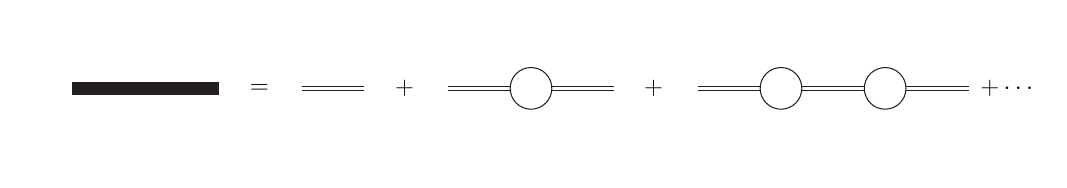}\\
  \caption{Full propagator for near-threshold states. The double line denotes the bare
  state.
  The particles in the loop are $D\bar{D}$.  }\label{ppgator}
  \end{center}
 \end{figure}

While Weinberg's original derivation relies on wave-function normalization and solving the Lippmann-Schwinger equation, we show here that the same relations can be obtained much more directly through on-shell renormalization in NREFT.
Hereafter, we adopt the convention in Refs.
\cite{Chen:2013upa,Xu:2024vne}, where the boson field operator
includes a factor of $\sqrt{2M}$ so that it has mass dimension
$3/2$.
Consider a bare state $|\mathcal{B}\rangle$ with bare mass $-B_0$,
width $\Gamma_0$, and coupling $g_0$ to the two-particle state
$D\bar{D}$. The interaction Lagrangian reads
\begin{equation}
\mathcal{L}_{int}=g_0\mathcal{B}D\bar D.
\end{equation}
We use $D\bar D$ to generically refer to two constituent particles that couple to the near-threshold state, with the masses $m_1$ and $m_2$, respectively.
The specific symbols are chosen for illustrative purposes, and the results derived here are completely general for
any two-particle system interacting in an s-wave.   
Near the two-particle threshold, the bare propagator
takes the form 
\begin{equation}
i\Delta_0=\dfrac{i}{E+B_0+i\Gamma_0/2}.\label{delta0}
\end{equation}
Note that the energy $E$ and the bare mass $-B_0$ are defined
relative to the $D\bar {D}$ threshold. Therefore, $B_0>0$ indicates
that the bare state lies below the threshold. $\Gamma_0$ denotes the
non-$D\bar {D}$ decay width of the bare state. We call
$|\mathcal{B}\rangle$ the bare state in the sense that it is the
eigenstate of $H_0$ rather than the total Hamiltonian $H$, where
$H=H_0+V$ with $V$ describing the interaction between
$|\mathcal{B}\rangle$ and $D\bar{D}$. The full propagator can be
obtained by summing the Feynman diagrams in Fig.\ref{ppgator}. Near
threshold, the particle momenta are non-relativistic. The loop
integral in Fig.\ref{ppgator} can be straightforwardly calculated
using the power divergence subtraction (PDS) scheme \cite{Kaplan:1998we},

\begin{align}
\mathcal{I}^{S}&\equiv(\Lambda/2)^{4-D}\int\frac{d^D\ell}{(2\pi)^D}\frac{1}{[\ell^0-\vec{\ell}^2/(2m_1)+i\epsilon]\cdot[E-\ell^0-\vec{\ell}^2/(2m_2)+i\epsilon]} \nonumber\\
&=-i(\Lambda/2)^{4-D}\int\frac{d^{D-1}\ell}{(2\pi)^{D-1}}\frac{1}{E-\frac{\vec{\ell}^2}{2\mu}+i\epsilon} \nonumber\\
&=i(\Lambda/2)^{4-D}\frac{2\mu}{(4\pi)^{\frac{D-1}{2}}}{(-2\mu E-i\epsilon)}^{\frac{D-3}{2}}\Gamma(\frac{3-D}{2}) \nonumber\\
&=-i\frac{\mu}{2\pi}(\sqrt{-2\mu E-i\epsilon}-\Lambda),\label{Intswave}
\end{align}
where $\Lambda$ is the renormalization scale introduced in the dimensional regularization, and the final result is obtained by subtracting the $D=3$ pole before taking the $D\to 4$ limit.
The full propagator then reads
\begin{equation}
i\Delta=\frac{i}{E+B_0-g_0^2\frac{\mu}{2\pi}(\sqrt{-2\mu{E}-i\epsilon}-\Lambda)+i\Gamma_0/2}.
\label{delta}
\end{equation}

\begin{figure}[t]
 \begin{center}
  \includegraphics[width=10cm]{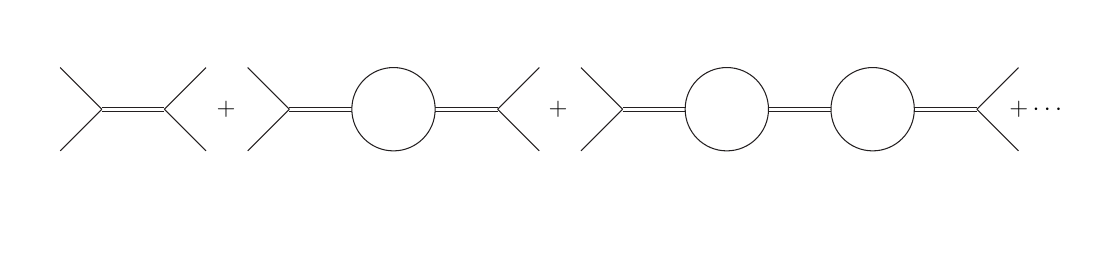}\\
  \caption{Feynman diagrams for two-particle scattering.}\label{amp}
  \end{center}
 \end{figure}

The Feynman diagrams for $D\bar{D}$ scattering are shown in
Fig.\ref{amp}, and the scattering amplitude is
\begin{equation}
\mathcal{A}=\frac{-g_0^2}{E+B_0-g_0^2\frac{\mu}{2\pi}(\sqrt{-2\mu{E}-i\epsilon}-\Lambda)+i\Gamma_0/2}.
\label{Aamp}
\end{equation}

We now demonstrate that, in the limit \(\Gamma_0 \to 0\), 
Weinberg's compositeness relations are equivalent to two on-shell renormalization conditions: (i) the propagator \(\Delta\) has a pole at \(E = -B\),
corresponding to a bound state. Here we use the term ``bound state’’
to mean a state lies below threshold, which should be distinguished
from a molecular state (corresponding to the special case $Z=0$ in
our language); and (ii) the residue of \(\Delta\) at this pole is
\(Z\).

Ref.~\cite{Chen:2026elc} discussed how to incorporate 
$\Gamma_0$ through coupled-channel effects, which allows for a unified treatment of bound states and resonances. In this section, however, our primary goal is to derive the compositeness relations. Therefore, we take the limit 
$\Gamma_0 \to 0$, following the original approach of Weinberg. This indicates that we first turn off the non-$D\bar D$ decay channel of the bare state $\mathcal{B}$, and fix the coupling strength $g_0^2$ using the on-shell renormalization conditions; we then turn on the non-$D\bar D$ decay channel in Sec.~\ref{Feynmanrules}. Turning on this decay channel does not affect the coupling strength $g_0^2$, as it contributes a purely imaginary part in the denominator of the propagator. In our scheme, $Z$ is defined as the residue of the propagator at the real-axis pole, after discarding the imaginary part of this denominator~\cite{Chen:2026elc}. In contrast, a different scheme defines $Z$ as the residue at the complex energy pole, which results a complex $Z$~\cite{Aceti:2012dd,Sekihara:2014kya,Kinugawa:2024crb}. 

Using the condition (i), we obtain
\begin{equation}
B_0=B+g_0^2\frac{\mu}{2\pi}(\sqrt{2\mu B}-\Lambda).\label{B01}
\end{equation}
With condition (ii), we obtain
\begin{equation}
g_0^2=\frac{2\pi\sqrt{2\mu B}}{\mu^2}\frac{1-Z}{Z}\label{g0}.
\end{equation}
Combining Eqs.\eqref{B01} and \eqref{g0}, we arrive at the relation first derived in Ref.~\cite{Chen:2013upa}
\begin{equation}
B_0=\frac{2-Z}{Z}B-\frac{1-Z}{Z}\sqrt{2B/\mu}\Lambda.\label{B02}
\end{equation}
Taking the limit \(\Gamma_0 \to 0\) and comparing Eq.\eqref{Aamp}
with the effective range expansion formula
\begin{equation}
\mathcal{A}=\frac{2\pi}{\mu}\frac{1}{-1/a+\frac{1}{2}r_0 p^2-ip},
\end{equation}
and noting that $E=p^2/2\mu$, we obtain
\begin{equation}
a=1/(\frac{2\pi B_0}{\mu g_0^2}+\Lambda),\ \ \ r_0=-\frac{2\pi}{\mu^2 g_0^2}.\label{ar2}
\end{equation}

Substituting Eqs.\eqref{g0} and \eqref{B02} into Eq.\eqref{ar2}, we
immediately recover Weinberg's compositeness relations in
Eq.\eqref{ar}. The result is clear: the low-energy scattering
amplitude has two independent parameters, which can be chosen as
($B,Z$) or ($a,r_0$). Therefore, two independent
relations are needed to connect these parameters. Condition (ii) i.e., Eq.\eqref{g0} is actually equivalent to the wave-function normalization condition used by Weinberg. This observation not only allows a natural extension of Weinberg's compositeness criteria to p-wave bound states (as presented here), but also paves the way for a unified treatment of resonances, as demonstrated in Ref.~\cite{Chen:2026elc}. While the wave-functions of resonances are non-normalizable, the condition (ii) remains applicable.

\section{Compositeness relations for near-threshold p-wave bound
states}\label{pwave}

We consider a bare state $\psi$ with bare mass $-B_{0}$, width
\(\Gamma_0\), and coupling constant \(g_0\) to the \(D\bar {D}\)
channel. Again, \(\Gamma_0\) denotes the non-\(D\bar {D}\) decay
width of $\psi$. Note that we use the same notation as in
Sec.\ref{swave} to amphasize that these are bare quantities. These
parameters should not be confused with those defined in
Sec.\ref{swave}. We first assume $\psi$ has spin one and couples to
the p-wave $D\bar {D}$ channel, where $D$ and $\bar D$ are spin-zero
particles. In Sec.\ref{Feynmanrules} we also discuss the case where
$\psi$ has spin zero. Near the $D\bar{D}$ threshold, the propagator
of this bare state can be written as

\begin{equation}
i\eta\delta^{ij}\Delta_0=\frac{
i\eta\delta^{ij}}{E+B_0+i\Gamma_0/2}.\label{PDelta0}
\end{equation}
We have included a factor $\eta$, as in Ref.
\cite{Bertulani:2002sz}, in the propagator, with $\eta=\pm 1$. More
about this will be discussed later. The p-wave scattering of
two-body systems in NREFT has been extensively studied since the
2000s \cite{Bertulani:2002sz,Bedaque:2003wa}. For convenience, we
use the interaction Lagrangian given in \cite{Chen:2012qq}
\begin{equation}
\mathcal{L}_{\psi D\bar D}=ig_{0}\{D^\dagger\nabla \bar D-\nabla D^\dagger\bar D\}\cdot\psi+
ig_{0}\{\bar D^\dagger\nabla D-\nabla \bar D^\dagger D\}\cdot\psi,\label{psiINT}
\end{equation}
where $D$($\bar D^\dagger$) annihilates a $D$($\bar D$) particle
while $D^\dagger$($\bar D$) creates a $D$($\bar D$) particle. With
this interaction Lagrangian, the full propagator shown in
Fig.\ref{ppgator} can be obtained using dimensional regularization
and the PDS scheme. The loop integral reduces to
\begin{align}
\mathcal{I}^{P}&\equiv(\Lambda/2)^{4-D}\int\frac{d^D\ell}{(2\pi)^D}\frac{\vec{\ell}^2}{[\ell^0-\vec{\ell}^2/2m_1+i\epsilon]\cdot[E-\ell^0-\vec{\ell}^2/2m_2+i\epsilon]}\nonumber\\
&=-i(\Lambda/2)^{4-D}\int\frac{d^{(D-1)}\ell}{(2\pi)^{(D-1)}}\frac{\vec{\ell}^2}{E-\vec{\ell}^2/2\mu+i\epsilon}\nonumber\\
&=i2\mu(2\mu E)(-2\mu
E-i\epsilon)^{(D-3)/2}\Gamma(\frac{3-D}{2})\frac{(\Lambda/2)^{4-D}}{(4\pi)^{(D-1)/2}}\nonumber\\
&=-i\frac{2\mu}{4\pi}(2\mu E)[(-2\mu E-i\epsilon)^{1/2}-\Lambda].\label{loopint}
\end{align}
The final result of the full propagator
reads
\begin{equation}
i\eta\delta^{ij}\Delta=\frac{i\eta\delta^{ij}}{E+B_0+i\Gamma_0/2-\frac{2\mu}{3\pi}\eta
g_0^2(2\mu E)[(-2\mu E-i\epsilon)^{1/2}-\Lambda]}.\label{PPGp}
\end{equation}
The
corresponding $D\bar D$ scattering amplitude, shown in Fig.
\ref{amp} reads
\begin{equation}
\mathcal{A}=\frac{-4\eta g_0^2
\vec{k}\cdot\vec{p}}{E+B_0+i\Gamma_0/2-\frac{2\mu}{3\pi}\eta
g_0^2(2\mu E)[(-2\mu E-i\epsilon)^{1/2}-\Lambda]},\label{PAmp}
\end{equation}
where $\vec{p}$ and $\vec{k}$ denote the incoming and outgoing
momenta of $D$, respectively. We now apply the two on-shell renormalization conditions stated in Sec.~\ref{swave}. First, we take the limit \(\Gamma_0 \to
0\) and set $E=-B$ as the bound state pole of $\Delta$. This gives
\begin{equation}
B_0=B-\frac{4}{3\pi}\eta g_0^2 \mu^2B(\sqrt{2\mu B}-\Lambda).\label{B0p}
\end{equation}
Second, letting the residue of $\Delta$ at this pole be $Z$, we
obtain
\begin{equation}
\eta g_0^2=\frac{3\pi}{4\mu^2}\frac{1-Z}{Z(\Lambda-\Lambda_0)},\label{g0p}
\end{equation}
where $\Lambda_0=\frac{3}{2}\sqrt{2\mu B}$.
Substituting Eq.\eqref{g0p} into Eq.\eqref{B0p} yields
\begin{equation}
B_0=\frac{\Lambda-(2+Z)\Lambda_0/3}{Z(\Lambda-\Lambda_0)}B.\label{B0p2}
\end{equation}
Taking the limit \(\Gamma_0\to 0\) and comparing the amplitude in
Eq.\eqref{PAmp} with the p-wave effective range expansion formula
\begin{equation}
\mathcal{A}=\frac{6\pi}{\mu}\frac{\vec{k}\cdot\vec{p}}{-\frac{1}{a_1}+\frac{1}{2}r_1
p^2-ip^3}.
\end{equation}
We obtain
\begin{equation}
a_1=\frac{2\mu\eta g_0^2}{3\pi B_0},\ \ \
r_1=-\frac{3\pi}{2\mu^2\eta g_0^2}-2\Lambda.\label{arp}
\end{equation} 
Substituting Eqs.\eqref{g0p} and \eqref{B0p2} into Eq.\eqref{arp}
gives the compositeness relations for p-wave bound states:
\begin{equation}
a_1=\frac{1-Z}{\Lambda-(2+Z)\Lambda_0/3}\frac{1}{2\mu B},\ \ \
r_1=-\frac{2}{1-Z}(\Lambda-Z\Lambda_0).\label{CRP}
\end{equation}

\begin{figure}[t]
 \begin{center}
  \includegraphics[width=10cm]{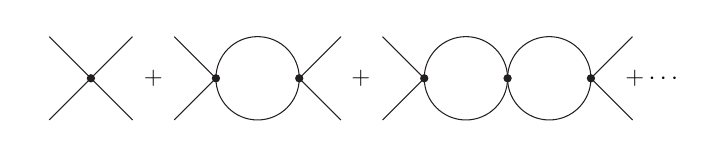}\\
  \caption{Feynman diagrams generated by the contact interaction.}\label{bubble}
  \end{center}
 \end{figure}

Unlike the s-wave case, the compositeness relations for p-wave states explicitly include the renormalization scale $\Lambda$.
However, the physical observables should be independent of $\Lambda$ due to the renormalization group running of $Z$.
Requiring both $a_1$ and $r_1$ to be independent of the renormalization scale $\Lambda$ leads to the same renormalization group equation for $Z$
\begin{equation}
\frac{dZ}{d\Lambda}=\frac{Z-1}{\Lambda-\Lambda_0}.\label{GEZ}
\end{equation}
This mutual consistency demonstrates that the compositeness relations are self-consistent within the NREFT framework. The scale dependence of $Z$ is a characteristic feature of the field-theoretic framework rather than a shortcoming. As we will demonstrate in the next section, it is this very dependence that enables us to attribute the negative-norm state to a specific choice of the subtraction scheme.

Before ending this section, we further discuss the NREFT framework
we have used. If we consider a $D\bar{D}$ p-wave interaction without
the bare state $\psi$, the interaction Lagrangian can be constructed
with contact interaction as
\begin{equation}
\mathcal{L}_{(D\bar D)^2}=C_2\{D^\dagger\nabla \bar
D-\nabla D^\dagger\bar D\}\cdot\{\nabla \bar D^\dagger
D-\bar D^\dagger\nabla D\}+\cdots,\label{CTINT}
\end{equation}
where we have omitted the higher-derivative terms. If a bound state
pole exists, one need to sum all the diagrams in Fig.\ref{bubble},
and the resulting amplitude reads
\begin{equation}
\mathcal{A}=\frac{4C_2\vec{k}\cdot\vec{p}}{1+\frac{2\mu}{3\pi}C_2(2\mu E)[(-2\mu
E-i\epsilon)^{1/2}-\Lambda]}.\label{BBAmp}
\end{equation}
Since $E=-B$ is the pole of the amplitude, we obtain
\begin{equation}
C_2=\frac{3\pi}{4\mu^2 B(\sqrt{2\mu B}-\Lambda)}.\label{C2}
\end{equation}
With this value and Eqs.\eqref{g0p}\eqref{B0p2}, one can easily check that, in the limit \(Z\to 0\),
the amplitude of the tree diagram in Fig.\ref{amp}, i.e., $\frac{-4\eta g_0^2 \vec{k}\cdot\vec{p}}{E+B_0+i\Gamma_0/2}$ reduces to the
amplitude of the tree diagram in Fig.\ref{bubble}, i.e., $4C_2 \vec{k}\cdot\vec{p}$. Also in the limit
\(Z\to 0\), the amplitude in Eq.\eqref{PAmp} reduces to that in
Eq.\eqref{BBAmp}. This is consistent with the equivalence between
the four-Fermi theory and the Yukawa theory found in
Ref.~\cite{Lurie}.

We now discuss the power counting of the NREFT. Near the threshold,
we treat the momentum $p$ and the binding scale $\gamma=\sqrt{2\mu B}$ (as well as $\Lambda_0=\frac{3}{2}\gamma$) as being of the same order,
scaling as $\mathcal{O}(p)$. An effective field theory makes use of the systematic expansion of $p/Q$, where $Q$ is the energy scale characterized by the heavier degrees of freedom which are not explicitly included in the effective field theory. To establish a consistent power counting in the NREFT, the renormalization scale 
$\Lambda$ must be taken to scale as $\mathcal{O}(p)$. Then $C_2$ from Eq.\eqref{C2} scales
$\mathcal{O}(p^{-3})$. One can check that all the diagrams in
Fig.\ref{bubble} are of the same order, i.e.,$\mathcal{O}(p^{-1})$,
and thus they need to be summed in the leading-order calculation,
and the resulting amplitude is also of order $\mathcal{O}(p^{-1})$.
This conclusion also applies to the summation of diagrams in
Fig.\ref{amp} and the amplitude in Eq.\eqref{PAmp}.

The tree-level amplitude in Fig.\ref{amp} is energy-independent for
$Z=0$, while for $Z\neq0$, it depends on energy as (with
\(\Gamma_0\to 0\))
\begin{equation}
\frac{1}{E+B_0}=\frac{1}{B_0}(1-x+x^2-x^3+\cdots), \ \ \
x\equiv\frac{p^2}{2\mu B_0}.\label{enexp}
\end{equation}
Eq.\eqref{B0p2} shows that $B_0$ is proportional to $B$ and scales as
$\mathcal{O}(p^2)$ for a non-negligible $Z$. Therefore, $x$ may not be treated as perturbatively small near the threshold. This observation
indicates that if $Z\neq 0$, the effect of the bare state cannot be mimicked by a finite number of contact terms. Thus, at $Z=0$, the interaction Lagrangian containing the $\psi$ field in Eq.\eqref{psiINT} is equivalent to a pure contact interaction in Eq.\eqref{CTINT}. In this case, the near-threshold bound state can be viewed as a pure composite state of $D\bar D$; whereas for $Z\neq 0$ the $\psi$ field must be included explicitly as an independent degree of freedom.

\section{Interpretation of negative-norm p-wave bound states}\label{negative}

More than a decade ago, Refs.~\cite{Nishida:2011np,Braaten:2011vf} pointed out that the $\psi$ field introduced in Eq.\eqref{psiINT} corresponds to a negative-norm state, which is unphysical. In this section, we
show that such a negative-norm state also emerges in the NREFT, but it can be transformed to a physical positive-norm state by an appropriate choice of the renormalization scale $\Lambda$.   

Let's return to Eq.\eqref{g0p}, 
if we chose $\Lambda<\Lambda_0$, for instance, we use the minimal subtraction (MS) scheme, which corresponds to setting $\Lambda=0$.
We then find that $\eta g_0^2$ is negative since $Z$ ranges from 0 to 1 according to the K\"all\'en-Lehmann representation~\cite{Kallen:1952zz,Lehmann:1954xi}. As $g_0^2$ is positive, we must set $\eta=-1$, which indicates the $\psi$ field corresponds to a negative-norm state. It is instructive to the study the nature of this negative-norm state by considering the normalization condition of
the wave function. Let $|G\rangle$ denote a
physical p-wave bound state with binding energy $B$. Its state vector can be written as
\begin{equation}
|G\rangle=\sqrt{Z}|\psi\rangle+\int d\alpha C_{\alpha}
|\alpha\rangle,\label{WF}
\end{equation}
where $|\psi\rangle$ is the bare state introduced in
Sec.\ref{pwave}, and $|\alpha\rangle$ denotes the continuum $D\bar
{D}$ state. The coefficient $C_\alpha$ can be expressed as
\begin{equation}
C_\alpha=\langle \alpha|G\rangle=\langle
\alpha|\frac{V}{H-H_0}|G\rangle=-\frac{\langle
\alpha|V|G\rangle}{E+B},
\end{equation}
where we have used $H|G\rangle=-B|G\rangle$ and
$H_0|\alpha\rangle=E|\alpha\rangle$. In the leading-order
calculation (or equivalently, neglecting the $D\bar D$ contact
interactions~\cite{Xu:2024vne}), Eq.\eqref{WF} gives
\begin{equation}
\langle \alpha|V|G\rangle=\sqrt{Z}\langle \alpha|V|\psi\rangle.
\end{equation}
To match the NREFT in Sec.\ref{pwave}, we use the following
parameterization
\begin{equation}
\langle \alpha |V|\psi\rangle=2g_0\vec{p}\cdot \vec{\epsilon},
\end{equation}
where $g_0$ is the coupling constant defined in Sec.\ref{pwave},
$\vec{p}$ is the momentum of $D$, and $\vec{\epsilon}$ is the
polarized vector of $|\psi\rangle$. With these notations, the wave
function normalization condition reads
\begin{equation}
\frac{4}{3}Zg_0^2\int\frac{d^3
p}{(2\pi)^3}\frac{p^2}{(p^2/2\mu+B)^2}=1-Z.\label{norm}
\end{equation}
Using dimensional regularization, the integral in Eq.\eqref{norm} is
evaluated as
\begin{eqnarray}
\mathcal{I^\prime}&\equiv&\int\frac{d^3
p}{(2\pi)^3}\frac{p^2}{(p^2/2\mu+B)^2}\nonumber\\
 &=&(\frac{\Lambda}{2})^{3-D}\int\frac{d^D
 p}{(2\pi)^D}\frac{p^2}{(p^2/2\mu+B)^2}.\nonumber\\
 &=&(\frac{\Lambda}{2})^{3-D}\frac{2D\mu^2}{(4\pi)^{D/2}}\Gamma(1-D/2)(2\mu
 B)^{D/2-1}.\label{Iprime}
\end{eqnarray}
Taking the limit $D\to 3$ in the MS scheme, we obtain
$\mathcal{I}^\prime=-\frac{3\mu^2}{2\pi}\sqrt{2\mu B}$. The
normalization condition in Eq.\eqref{norm} then yields
\begin{equation}
g_0^2=\frac{\pi}{2\mu^2\sqrt{2\mu B}}\frac{Z-1}{Z}.
\end{equation}
This matches the result in Eq.\eqref{g0p} with $\eta=1$, and $ \Lambda=0$, yielding a
negative $g_0^2$. The origin of the issue is clear: the integral
$\mathcal{I}^\prime$ is positive by definition, but the MS scheme
subtracts too much, rendering it negative. To resolve this, we 
adopt the PDS scheme, which remove the $D=2$ pole in
Eq.\eqref{Iprime}. This gives
\begin{equation}
\mathcal{I}^\prime=\frac{\mu^2}{\pi}(\Lambda-\Lambda_0).\label{Ippp}
\end{equation}
For $\Lambda>\Lambda_0$, $\mathcal{I}^\prime$ is
positive. Substituting Eq.\eqref{Ippp} into Eq.\eqref{norm} yields
\begin{equation}
g_0^2=\frac{3\pi}{4\mu^2}\frac{1-Z}{Z(\Lambda-\Lambda_0)}.\label{g0pp}
\end{equation}
This exactly reproduces the result in Eq.\eqref{g0p} with
$\eta=1$. To ensure that $g_0^2$ is positive, $\Lambda$ must satisfy the constraint $\Lambda>\Lambda_0$. 

We conclude this section by emphasizing that the negative-norm state can be avoided through an appropriate choice of
the renormalization scale $\Lambda$, specifically, $\Lambda>\Lambda_0$.
In standard renormalization theory, the scale $\Lambda$ is arbitrary, provided it is of the order of the typical momentum scale of the concerned energy regime. However, we find here that the expression for $g_0^2$ i.e., Eq.\eqref{g0pp}, has a singular point at $\Lambda=\Lambda_0$. On the one side, ($\Lambda>\Lambda_0$), the near-threshold state is a positive-norm state, while on the other side ($\Lambda<\Lambda_0$), it turns into a negative-norm state. Certainly, the physical near-threshold p-wave bound state should be a positive-norm state; thus we should choose $\Lambda>\Lambda_0$ in practical calculations instead of $\Lambda=0$ (the MS scheme). This choice is necessary because the singular point prohibits one from performing the analytical continuation from $\Lambda>\Lambda_0$ to $\Lambda<\Lambda_0$.

\section{The general propagator for p-wave near-threshold states 
}\label{Feynmanrules}

Since near-threshold p-wave bound states are of the comment interest in hadron physics, nuclear physics and ultracold atom physics, our formalism is expected to have broad applications. For example, if the two-body elastic scattering data are available, the compositeness relations presented in this work can used to extract the renormalization constant $Z$, which provides information on the nature of these near-threshold states. If such data are unavailable or if the the near-threshold states have a non-negligible width (which is usually the case in hadron physics) one can use a general propagator to fit the lineshape data and extract the value of $Z$. An Example of such a study is given in Refs.~\cite{Chen:2013upa,Xu:2024vne,Xu:2023lll}, where the near-threshold state is the famous $X(3872)$, an s-wave near-threshold state. The propagator for the near-threshold state given in Ref.~\cite{Xu:2024vne} is general in the sense that it explictly contains the factor $Z$. It reduces to Breit-Wigner form at $Z=1$, reduces to the formula for a pure molecular state in Ref.~\cite{Braaten:2007dw} at $Z=0$, and is equivalent to the Flatt$\acute{e}$ form~\cite{Flatte:1976xu} for $0<Z<1$.
 
Here, we present the general
propagator for spin-one near-threshold p-wave bound states, which reads

\begin{eqnarray}
G_1(E)&=&\frac{iZ \delta^{ij}}{D_{EFT}(E)},\nonumber\\
D_{EFT}(E)&=& E+B+\Sigma(E)+i\Gamma/2,\nonumber\\
\Sigma(E)&=&g^2\frac{2\mu}{3\pi}[(-2\mu
E-i\epsilon)^{3/2}+\mu\sqrt{2\mu B}(3E+B)],\label{G1}
\end{eqnarray}
where
\begin{equation}
\Gamma=Z\Gamma_0,\ \ \ g^2\equiv
Zg_0^2=\frac{3\pi}{4\mu^2}\frac{1-Z}{\Lambda-\Lambda_0}.\label{egg0}
\end{equation}
Requiring the renormalized coupling constant $g^2$ to be independent of the renormalization scale $\Lambda$ leads to exactly the same equation as in Eq.\eqref{GEZ}, and its solution reads
\begin{equation}
Z=1-C(\Lambda-\Lambda_0).\label{SREGZ}
\end{equation}
For $Z$ ranges from 0 to 1, the coefficient $C$ must satisfy $0<C<\frac{1}{\Lambda-\Lambda_0}$. Substituting Eq.\eqref{SREGZ} into Eq.\eqref{egg0}, we obtain
\begin{equation}
g^2=\frac{3\pi}{4\mu^2}C,
\end{equation}
which is explicitly independent of the renormalization scale $\Lambda$. The coefficient $C$ can be extracted by fitting the experimental lineshape with the propagator.

\begin{figure}[t]
 \begin{center}
  \includegraphics[width=10cm]{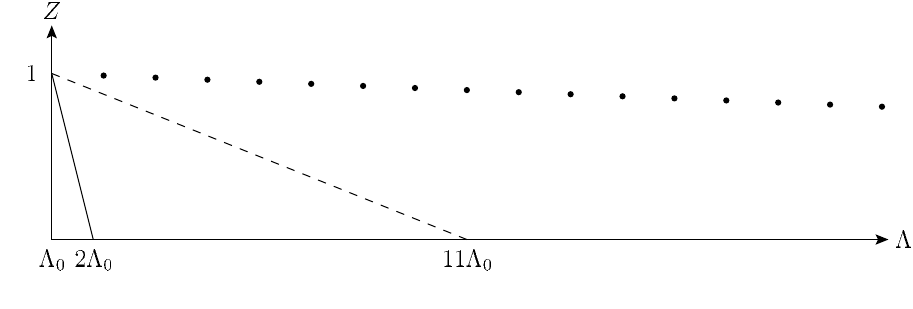}\\
  \caption{Running of $Z$ with respect to $\Lambda$ for different values of $C$:$C=1/\Lambda_0$(solid line), $C=1/(10\Lambda_0)$(dashed line), and $C=1/(100\Lambda_0)$(dotted line).}\label{runz}
  \end{center}
 \end{figure}

The dependence of $Z$ on $\Lambda$ may raise concerns that $Z$
could run from 1 to 0 as $\Lambda$ varies, thereby undermining its usefulness for distinguishing the structure of near-threshold states. This concern is not necessarily warranted. The running of $Z$ depends on the magnitude of $C$ (see Fig.\ref{runz} for illustration). As mentioned previously, an effective field theory makes use of the expansion of $p/Q$, where $Q\gg p$ and $\Lambda$---as well as $\Lambda_0$---scales as $\mathcal{O}(p)$. If $C$ scales as $\mathcal{O}(p^{-1})$, $Z$ will exhibit a rapid running with $\Lambda$; conversely, if $C$ scales as $\mathcal{O}(Q^{-1})$, the running of $Z$ with $\Lambda$ will be much milder. For example, if $C=1/(100\Lambda_0)$, $Z$ would run to zero at $\Lambda=101\Lambda_0$; however this scale is beyond the valid regime of the effective field theory. The value of $Z$ should be extracted at a fixing $\Lambda$ that is of order $\mathcal{O}(p)$ but chosen to be well separated from $\Lambda_0$. This is because at $\Lambda=\Lambda_0$ the second term on the right-hand side of Eq.\eqref{SREGZ} vanishes regardless of the magnitude of $C$, thereby losing the information about the nature of the near-threshold state. If the extracted $Z$ is close to 1, it implies that
$C$ is small, and the $\Lambda$-dependence of $Z$ is weak, indicating that the bare-state component in the physical bound state $|G\rangle$ is dominant. Conversely, if $Z$ is close to 0, $C$ is not small, and $Z$ becomes more sensitive to $\Lambda$, implying a negligible bare-state component. In other words, if $Z$ depends sensitively on $\Lambda$, the effect of the bare state can be removed by slightly changing the value of $\Lambda$, making it unnecessary to include $\psi$ in the Lagrangian explicitly. This indicates that the near-threshold state is a pure molecular state. If $Z$ does not depend sensitively on $\Lambda$, it indicates that $\psi$ should be included explicitly in the Lagrangian and the near-threshold state contains a substantial bare-state component. Therefore, the value of $Z$ can still be used to infer the internal structure of near-threshold states.

It is worth noting that Eq.\eqref{G1} is the reformulated form of Eq.\eqref{PPGp}
after substituting Eqs.\eqref{g0p} and \eqref{B0p2}. The propagator
in Eq.\eqref{PPGp} is of the Flatt$\acute{e}$
form~\cite{Flatte:1976xu}, whose parameters ($B_0$,$g_0^2$) become
infinite in the limit \(Z\to 0\), as is evident from Eqs.
\eqref{g0p} and \eqref{B0p2}. We therefore recommend using
Eq.\eqref{G1} to fit the lineshape data, since $D_{EFT}(E)$ remains
well-defined in the limit \(Z\to 0\). This is particularly important
because most near-threshold exotic states are commonly identified as
molecular states, i.e., $Z=0$. One can also readily see that for
$Z=1$~($g^2=0$), Eq.\eqref{G1} reduces to the Breit-Wigner
form. Hence Eq.\eqref{G1} serves as a general
propagator for near-threshold p-wave states. The Feynman vertex for
the coupling between $|G\rangle$ and the $|D\bar D\rangle$ channel
is $-2ig_0 p^{i}$, where $g_0$ is given in Eq.\eqref{egg0}, and $p$
is the momentum of particle $D$.

For a spin-zero near-threshold state $|G_0\rangle$ coupled to the
p-wave $|D^\ast\bar D\rangle$ channel, where $D^\ast$ is a spin-one
particle and $\bar D$ is spin-zero, the general propagator is
\begin{equation}
G_0(E)=\frac{iZ}{D_{EFT}(E)},
\end{equation}
with the same $D_{EFT}(E)$ as in Eq.\eqref{G1}. The
vertex for the coupling between $|G_0\rangle$ and $|D^\ast\bar
D\rangle$ is $-\frac{2}{\sqrt{3}}ig_0 p^{i}$, with $g_0$ given by
Eq.\eqref{egg0}.

In this paper, only one continuum channel, $D\bar D$, is considered. For the newly observed state $G(3900)$, both the neutral and charged channels of $D\bar D$ must be taken into account. In this case, the coupled-channel effect has to be treated, which is beyond the scope of this paper and is discussed in Ref.~\cite{Chen:2026elc}.

\section{Summary}\label{summary}

We reformulate Weinberg's compositeness relations as two simple on-shell renormalization 
conditions within the NREFT framework. These conditions can be
straightforwardly applied to study near-threshold p-wave bound
states, for which we derive the corresponding compositeness
relations. We pointed out that the renormalization scale $\Lambda$ in the effective field theory should be chosen as $\Lambda>\Lambda_0$ to avoid the negative-norm state in p-wave sector. We then provide the general propagator for p-wave near-threshold states, which can be used to fit the experimental lineshape data and study the nature of the near-threshold states. Since near-threshold p-wave bound states are of common interest in hadron physics, nuclear physics, and ultracold atomic physics, our results are expected to have broad applications. Finally, we have
considered only the leading-order effects of the NREFT. To go beyond the leading order, one can further include contact interactions and t-channel meson-exchange interactions.

\acknowledgments

The author used DeepSeek for literature discovery and language editing.




  
%

\end{document}